\documentclass[sigconf]{acmart}

\acmConference[ICSE 2024]{46th International Conference on Software Engineering}{April 2024}{Lisbon, Portugal}

\usepackage[breakable,skins]{tcolorbox}
\usepackage{framed}
\usepackage{multirow, makecell}
\usepackage{lipsum}  
\usepackage{url} 
\usepackage{xspace}
\usepackage{graphicx} 
\usepackage{hyperref}

\newcommand{\emo}[1]{{\em {\MakeUppercase #1}}}
\newcommand{\zerollm}{zero-shot LLMs\xspace}
\newcommand{\ece}{emotion-cause extraction\xspace}

\newcommand{\revisiontext}[1]{#1}

\title{Uncovering the Causes of Emotions in Software Developer Communication Using Zero-shot LLMs}
\author{Mia Mohammad Imran}
\affiliation{\institution{Virginia Commonwealth University}
  \city{Richmond, Virginia}
  \country{USA}
}
\email{imranm3@vcu.edu}

\author{Preetha Chatterjee}
\affiliation{\institution{Drexel University}
  \city{Philadelphia, Pennsylvania}
  \country{USA}
}
\email{pc697@drexel.edu}

\author{Kostadin Damevski}
\affiliation{\institution{Virginia Commonwealth University}
  \city{Richmond, Virginia}
  \country{USA}
}
\email{kdamevski@vcu.edu}

\begin{document}

\begin{abstract}
Understanding and identifying the causes behind developers' emotions (e.g., \emo{Frustration} caused by `delays in merging pull requests') can be crucial towards finding solutions to problems and fostering collaboration in open-source communities. Effectively identifying such information in the high volume of communications across the different project channels, such as chats, emails, and issue comments, requires automated recognition of emotions and their causes. To enable this automation, large-scale software engineering-specific datasets that can be used to train accurate machine learning models are required. However, such datasets are expensive to create with the variety and informal nature of software projects' communication channels.

In this paper, we explore zero-shot LLMs that are pre-trained on massive datasets but without being fine-tuned specifically for the task of detecting emotion causes in software engineering: ChatGPT, GPT-4, and flan-alpaca. \revisiontext{Our evaluation indicates that these recently available models can identify emotion categories when given detailed emotions, although they perform  worse than the top-rated models. For emotion cause identification, our results indicate that zero-shot LLMs are effective at recognizing the correct emotion cause with a BLEU-2 score of 0.598.} To highlight the potential use of these techniques, we conduct a case study of the causes of \emo{frustration} in the last year of development of a popular open-source project, revealing several interesting insights.
\end{abstract}

\maketitle

\section{Introduction}

Emotions play a crucial role in software engineering, influencing individual and team performance, communication, and decision-making. Numerous software engineering tasks have been found to be impacted by developer emotions, e.g., bug fixing efficiency~\cite{Destefanis2016SoftwareDD, ortu2015bullies}, build success of continuous integration~\cite{7962396}. Previous research has also studied the link between developers' emotions, their productivity, and their attrition in software development projects~\cite{Murgia2014DoDF, Graziotin13, madampe2023framework}. Open source projects' success depends on attracting and retaining volunteer participants. Automatically detecting the emotions in the communication channels of a software project, e.g., pull requests and issues comments, chats, and discussion boards, can provide valuable insights into improving the outcomes of open-source software projects. 

Going beyond detecting the occurrence of different emotions in developer communication channels, identifying the causes of those emotions is key for many uses. Simply knowing the existence of an emotion is often insufficient for understanding what or whom the expressed emotion is towards and for determining the appropriate reaction~\cite{gachechiladze2017anger}. However, when the emotion cause is known, along with the type of emotion, it becomes possible to reliably assess potential implications. This could allow for understanding developer opinion towards different aspects of the project like technical debt~\cite{Fucci21} or code reviews~\cite{Ebert2021AnES}. For instance, a developer made the following comment on an open issue in the {\sf flutter/flutter} GitHub project, {\em ``this is a really severe issue, the ux is pretty awful when you have a splash and then a landing page to simulate splash because it is very obvious that is a different view than the splash"}, which expresses \emo{frustration} (a sub-emotion of \emo{anger}). The cause of this emotion is the text span, {\em ``the ux is pretty awful"}. Extracting emotion causes automatically is challenging because of the distinct nature of software engineering communication (e.g., it includes domain-specific idioms like {\em `spaghetti code'}), the variety of different channels (e.g., chats vs. issue comments), and the informal nature of developer communication (e.g., often containing informal abbreviations like `AFAIK'~\cite{Chatterjee21}). It is likely that \ece requires a large amount of software engineering-specific training data that can capture this variability, in both emotion and language~\cite{yu2019multiple, xu2019extracting}.

Large Language Models (LLMs) have recently emerged as a new powerful type of deep learning technique. These models are built by unsupervised pre-training on a very large dataset, followed by supervised fine-tuning on a smaller dataset. During pre-training, the model learns to predict the next word, given a sequence of words. During fine-tuning, the model is provided with labeled data relevant to a specific task. Some of the largest and most powerful LLMs, such as ChatGPT~\cite{chatgpt} and GPT-4~\cite{OpenAI2023GPT4TR}, are now widely available but do not disclose details about their dataset, training process, or model weights. Consequently, fine-tuning them for a specific task or dataset, such as detecting emotion-causes in software engineering text, is not possible. However, these LLMs can still be used as ``zero-shot" models, where no task-specific fine-tuning is performed. Since constructing a large training dataset for \ece task in software engineering communication is expensive, using a zero-shot setup is an attractive option. 
Recently, models used as "zero-shot" setup, has been often refer as "zero-shot LLMs" in literature~\cite{carta2023iterative, zhuang2023beyond, wang2023zero, kojima2022large}. In this research, we have used same phrase to refer when a model has been used in zero-shot setup.

This paper applies 3 models in zero-shot setup, ChatGPT~\cite{chatgpt}, GPT-4~\cite{OpenAI2023GPT4TR}, and one that is open-source, flan-alpaca~\cite{flan-alpaca}, to the problem of \ece in software engineering. We first examine the ability of such models to detect emotions in software engineering text, relative to state-of-the-art techniques and to LLMs fine-tuned for detecting emotions. 
Next, we examine the effectiveness of the \zerollm  for the \ece task. The results indicate that these models are promising, achieving a BLUE-2 score of 0.598 on a manually curated dataset of 450 utterances. Finally, we perform a case study on the causes of \emo{Frustration}, an undesirable emotion within a large open-source software project~\cite{ford2015exploring}, to further highlight the utility of \ece for software engineering. 
\noindent
The main contributions of this paper are:
\begin{itemize}
    \item application and evaluation of \zerollm to the problem of \ece in software engineering.
    \item manually-curated \ece dataset of 450 GitHub comments.
    \item case study highlighting the usefulness and purpose of automatic \ece in software engineering.
    \item evaluation of \zerollm compared to state-of-the-art techniques, including fine-tuned LLMs, on the well-known problem of classifying emotions in software engineering communication.
\end{itemize}

We publish the source code and annotated dataset to facilitate the replication of our study at:  \url{https://anonymous.4open.science/r/SE-Emotion-Cause-Replication-0C01}.
 \section{Preliminary Study: Detecting Emotion Types}
\label{prelim_study}

Detecting the causes of emotions in text requires a reliable model that can accurately identify the type of emotion expressed. Therefore, before proceeding, we conduct a preliminary investigation to determine if \zerollm can accurately detect emotions in software engineering texts. We compare the performance of these models with 1) existing state-of-the-art emotion classification models in software engineering, and 2) fine-tuned LLMs. 
The models are evaluated on three different types of datasets: a) GitHub comments dataset~\cite{imran2022data}, b) Stack Overflow comments dataset~\cite{novielli2018gold}, and c) JIRA comments dataset~\cite{ortu2015bullies}.

\subsection{Datasets}

\textbf{GitHub Dataset}. Imran et al. curated a diverse collection of 2000 data points sourced from GitHub issues and pull requests comments~\cite{imran2022data}. The dataset is manually annotated with six distinct emotion classes: \emo{Anger}, \emo{Love}, \emo{Fear}, \emo{Joy}, \emo{Sadness}, and \emo{Surprise}. Among the comments, 17\% convey \emo{Anger}, 11\% \emo{Love}, 9.90\% \emo{Fear}, 21.10\% \emo{Joy}, 13.70\% \emo{Sadness}, and 16.40\% \emo{Surprise}. The remaining comments remain devoid of associated emotions.

\noindent
\textbf{Stack Overflow Dataset}. Novielli et al. annotated a rich multi-label dataset comprising 4800 Stack Overflow questions, answers, and comments~\cite{novielli2018gold}. Within this dataset, 18.1\% of the samples are labeled with \emo{Anger}, 25.4\% with \emo{Love}, 2.2\% with \emo{Fear}, 10.2\% with \emo{Joy}, 4.8\% with \emo{Sadness}, and 0.9\% with \emo{Surprise}. The remaining contents of the dataset are neutral.

\noindent\textbf{JIRA Dataset}. Ortu et al. annotated a comprehensive collection of 4000 comments extracted from JIRA, classifying them into four distinct emotional categories: \emo{Love}, \emo{Joy}, \emo{Anger}, and \emo{Sadness} (1000 comments each)~\cite{ortu2015bullies}. Within each category, \emo{Love}, \emo{Joy}, \emo{Sadness}, and \emo{Anger} account for 16.6\%, 12.4\%, 32.4\%, and 30.2\% respectively, while the remaining comments are neutral.

For training and testing with each dataset we employ an 80\%-20\% stratified sampling approach.

\subsection{Emotion Model}

All these three datasets rely on the well-known Shaver's tree-structured emotion model~\cite{Shaver}. In Shaver's model, for each of the six basic emotions, there are secondary and tertiary-level emotions, which refine the granularity of the previous level. GoEmotions is an alternative emotion model used in the literature that was proposed by researchers at Google with the focus on emotions that can be observed in written text~\cite{goemotions}. In their recent work, Imran et al.~\cite{imran2022data} extended Shaver's model by incorporating a few emotions from GoEmotions's~\cite{goemotions} taxonomy in order to study emotions present in GitHub communications. Out of 27 emotions in GoEmotions' list, 26 are in the extended model by Imran et al. The only emotion that is not on the list is \emo{Gratitude}. 

In order to study both GoEmotions and Shaver's models, in this paper, we map the remaining emotion - \emo{Gratitude} - within Shaver's tree-structured emotion model. We look into the definitions - how the authors defined \emo{Gratitude} in GoEmotions~\cite{goemotions} and if any emotion is defined similar way in Shaver et al.~\cite{Shaver}'s definition. GoEmotions defined \emo{Gratitude} as \textit{``a feeling of thankfulness and appreciation."}, while Shaver et al. defined \emo{Love} \textit{``involving the appreciation of someone."} Therefore, we mapped \emo{Gratitude} as a secondary emotion to the basic emotion \emo{Love} in this study.

The extended model is shown in Table~\ref{tab:shavers_category}, with blue-colored emotions also appearing in the GoEmotions' listing.

\begin{table}
\begin{center}
\small
\caption{Extended Shaver's tree-structured taxonomy.}
\begin{tabular} { | l | p{6.5cm} | }
\hline
    Basic Emotion & {Secondary Emotion $\rightarrow$} Tertiary Emotion\\
    \hline 
    
     \multirowcell{10}{ Anger } & {Irritation $\rightarrow$}  \textcolor{blue}{Annoyance}, Agitation, Grumpiness, Aggravation, Grouchiness \\
     & {Exasperation $\rightarrow$}  Frustration \\
     & {Rage $\rightarrow$}  \textcolor{blue}{Anger}, Fury, Hate, Dislike, Resentment, Outrage, Wrath, Hostility, Bitterness, Ferocity, Loathing, Scorn, Spite, Vengefulness \\
     & {Envy $\rightarrow$}  Jealousy \\
     & {\textcolor{blue}{Disgust} $\rightarrow$}  Revulsion, Contempt, Loathing \\
     & {Torment}  \\ 
     & {\textcolor{blue}{Disapproval}} $\dagger$ \\
     \hline

     \multirowcell{6}{Love } & {Affection $\rightarrow$}  Liking, \textcolor{blue}{Caring}, Compassion, Fondness, Affection, \textcolor{blue}{Love}, Attraction, Tenderness, Sentimentality, Adoration \\
     & {Lust $\rightarrow$} \textcolor{blue}{Desire}, Passion, Infatuation \\
     & Longing \\ 
     & \textcolor{blue}{Gratitude} $\ddagger$ \\
    \hline
    
     \multirowcell{4}{ Fear } & {Horror $\rightarrow$}  Alarm, Fright, Panic, Terror, \textcolor{blue}{Fear}, Hysteria, Shock, Mortification \\
     & {\textcolor{blue}{Nervousness} $\rightarrow$} Anxiety, Distress, Worry, Uneasiness, Tenseness, Apprehension, Dread \\
     \hline
     
     \multirowcell{11}{Joy}
     & {Cheerfulness $\rightarrow$} Happiness, {\textcolor{blue}{Amusement}}, Satisfaction, Bliss, Gaiety, Glee, Jolliness, Joviality, {\textcolor{blue}{Joy}}, Delight, Enjoyment, Gladness, Jubilation, Elation, Ecstasy, Euphoria \\
     & {Zest $\rightarrow$} Enthusiasm, {\textcolor{blue}{Excitement}}, Thrill, Zeal, Exhilaration \\
     & {Contentment $\rightarrow$} Pleasure \\
     & {{\textcolor{blue}{Optimism}} $\rightarrow$} Eagerness, Hope \\
     & {{\textcolor{blue}{Pride}} $\rightarrow$} Triumph \\
     & {Enthrallment $\rightarrow$} Enthrallment, Rapture \\
     & {\textcolor{blue}{Relief}}\\
     & {\textcolor{blue}{Approval}} $\dagger$ \\
     & {\textcolor{blue}{Admiration}} $\dagger$ \\
     \hline

    \multirowcell{10}{ Sadness} & {Suffering $\rightarrow$} Hurt, Anguish, Agony \\
    & {{\textcolor{blue}{Sadness}} $\rightarrow$} Depression, Sorrow, Despair, Gloom, Hopelessness, Glumness, Unhappiness, {\textcolor{blue}{Grief}}, Woe, Misery, Melancholy \\
    & {{\textcolor{blue}{Disappointment}} $\rightarrow$} Displeasure, Dismay \\
    & {Shame $\rightarrow$} Guilt, Regret, {\textcolor{blue}{Remorse}} \\
    & {Neglect $\rightarrow$} {\textcolor{blue}{Embarrassment}}, Insecurity, Insult, Rejection, Alienation, Isolation, Loneliness, Homesickness, Defeat, Dejection, Humiliation \\
    & {Sympathy $\rightarrow$} Pity \\
    \hline
    \multirowcell{4}{ Surprise} & {\textcolor{blue}{Surprise} $\rightarrow$} Amazement, Astonishment \\
    & \textcolor{blue}{Confusion} $\dagger$  \\
    & \textcolor{blue}{Curiosity} $\dagger$ \\
    & \textcolor{blue}{Realization} $\dagger$ \\
    \hline 
\end{tabular}
\label{tab:shavers_category}
\end{center}
\footnotesize
\raggedright 
{\bf Notes:} Emotions in blue appear in the list of emotions proposed by GoEmotions. Emotions added by Imran et al. from GoEmotions' list onto Shaver's taxonomy are denoted with $\dagger$. A single emotion -- \emo{Gratitude} -- is added to the taxonomy by this paper, denoted by $\ddagger$.
\end{table}

\subsection{Compared Models}

\noindent
\textbf{Existing SE-specific models.} We use three existing SE-specific models that have been shown to produce state-of-the-art performance in emotion classification.

\begin{itemize}
    \item \textit{ESEM-E}: A tool that is proposed by Murgia et al.~\cite{esem-e} that uses unigram and bigram as features and an SVM as the ML model. It has been widely used in the literature for software engineering emotion classification tasks~\cite{esem-e, sentimoji, imran2022data}.

    \item \textit{EMTk}: EMTk is proposed by Calefato et al.~\cite{emtk}. EMTk uses unigram, bigram, emotion lexicon, politeness, and mood as features and SVM as the ML model. Similar to ESEM-E, it has been widely used in the SE community~\cite{novielli2018gold, sentimoji, imran2022data}.
    
    \item \textit{SEntiMoji}: This deep learning-based model is proposed by Chen et al.~\cite{sentimoji}, and built on top of the DeepMoji~\cite{deepmoji} model. This flexible model can identify different emotion categorization schemes, including Shaver’s categorization.

\end{itemize}

For EMTk and SEntiMoji, the authors published the model implementations. We use the provided code for training and testing. As for ESEM-E, we carefully read the instructions provided by the authors and implemented the model by ourselves.

\noindent
\textbf{Fine-tuned LLMs.} We fine-tune two popular LLMs -- BERT and RoBERTa -- that have been widely used as emotion and sentiment analysis, including in software engineering~\cite{batra2021bert, eeshita-sentiment, kamath2022emoroberta, liu2019dens}. We leverage the pre-trained model weights from HuggingFace~\cite{wolf2019huggingface}.

\begin{itemize}
    \item \textit{BERT}: Bidirectional Encoder Representations from Transformers is a widely-used LLM developed by Google. BERT is pre-trained using English Wikipedia and BooksCorpus~\cite{bert}.
    \item \textit{RoBERTa}: Robustly Optimized BERT, a variant of BERT, is developed by Meta. It is pre-trained using English Wikipedia, BooksCorpus, news articles, Web text, and stories~\cite{roberta}.
\end{itemize}

\noindent
\textbf{Zero-shot LLMs.} We use three (two commercial and one open-source) recent pre-trained and instruction-tuned models in a zero-shot setting, i.e., the models are not tuned for the task of emotion (cause) detection in software engineering. 

\begin{itemize}
    \item \textit{ChatGPT (GPT-3.5-turbo)}~\cite{chatgpt}: We use the {\sf gpt-3.5-turbo} API by OpenAI. GPT-3.5-based models are pre-trained on a massive corpus of text data from diverse sources, including books, articles, websites, and other publicly available online content. The model was then instruction-tuned (from a large dataset of instructions with desired output) using Reinforcement Learning from Human Feedback (RLHF)~\cite{Ouyang2022TrainingLM}.
    
    \item \textit{GPT-4}~\cite{OpenAI2023GPT4TR}: We use the {\sf gpt-4} API by OpenAI. GPT-4 is a transformer-style model pre-trained using both publicly available data and data licensed from third-party providers; details of the training data are not released at the time of writing. GPT-4 introduced a rule-based reward model (RBRM) approach on top of RLHF.
    
    \item \textit{flan-alpaca}~\cite{flan-alpaca}: This is a variation of the Alpaca~\cite{alpaca} fine-tuned model. Alpaca was developed by Stanford, based on Meta's LLaMA~\cite{llama} model using 52K instruction-based data instances. Due to licensing issues, the original Alpaca model is not accessible at the time of our experiment. Instead, using the Alpaca instructions dataset, Chia et al.~\cite{flan-alpaca} fine-tuned Google's instruction-tuned Flan-T5~\cite{flan} model and released the weights. We use the {\sf flan-alpaca-xl} version from Hugging Face~\cite{flan-alpaca-xl}.
\end{itemize}

\subsection{Metrics}

The F1-score is a widely used metric for assessing the effectiveness of a (multi-class) classification model. It is the weighted harmonic mean of precision and recall, which takes into account both false positives and false negatives. $F1-score = 2*\frac{Recall*Precision}{Recall + Precision}$. To calculate the average score across all classes, i.e., emotions, we use the micro-averaged variant which has been widely used in related tasks~\cite{liu2019dens, imran2022data, sokolova2009systematic}.

\subsection{Basic Emotion Prompting}

The \zerollm we are considering are all instruction- (or prompt-) tuned. This recent category of LLMs use a fine-tuning process with instructional data, which helps the LLMs to better comprehend and respond to user-composed prompts. To our knowledge, there is no prior work on how to formulate prompts for emotion recognition in software engineering text using these LLMs.

A recent study by Kocon et al.~\cite{kocon2023chatgpt} evaluated the performance of ChatGPT on various natural language processing tasks by designing over 38k prompts that covered 25 different tasks, including emotion classification using the GoEmotions dataset\cite{goemotions}. Inspired by this study, we designed a prompt for emotion classification that we used on all three datasets. More specifically, we asked the models to act as a user in a specific platform, i.e., GitHub, Stack Overflow, and JIRA, and provided the utterances and a list of the basic (top-level) emotions: \emo{Anger}, \emo{Fear}, \emo{Love}, \emo{Joy}, \emo{Sadness}, and, \emo{Surprise}. The prompt is the following: 

\begin{tcolorbox}[breakable,colback=black!03!white]
{\sf \noindent
You are a [GitHub/Stack Overflow/JIRA] user. You are reading comments from [GitHub/Stack Overflow/JIRA].
Your task is to detect whether there is one of the following emotions aroused in you while reading the utterance.

\noindent
Emotions List: Anger, Fear, Love, Joy, Sadness, Surprise.

\smallskip

\noindent
Utterance: \textcolor{blue}{\em <insert utterance>}.

\smallskip

\noindent
If there is no emotion in the text, write Neutral. Otherwise write exactly one word, the exact emotion from the emotions list.
}
\end{tcolorbox}

\noindent
Since the JIRA dataset does not contain \emo{Fear} and \emo{Surprise}, we do not list these two emotions in the prompt when evaluating with this dataset.

\noindent
{\bf Results and Discussion.} Table~\ref{tab:classification_summary} shows the results for the three emotion classification datasets and for all the models. 
It is clearly noticeable from the results that the \zerollm performed poorly across all datasets, lagging behind the SE-specific models and the fine-tuned LLM models. The fine-tuned LLMs performed best, e.g., RoBERTa achieved the best micro-averaged F1-score overall by averaging 0.592, 0.735, and 0.818 respectively for GitHub, Stack Overflow, and JIRA datasets. Among the SE-specific models, the deep learning-based SEntiMoji model performed best with an average F1-score of 0.529. 

In order to understand where the \zerollm are making mistakes, next, we conduct an error analysis.

\smallskip
\noindent
\textbf{Error analysis.} One of the most common errors we observed is that \zerollm are misclassifying \emo{Love} utterances as \emo{Joy} for all datasets. For example, on the Stack Overflow dataset, the F1-score for \emo{Love} is 0.0, 0.116, and 0.078 for flan-alpaca, ChatGPT, and GPT-4 respectively. Compared to this, BERT, RoBERTa, ESEM-E, EMTk, and SEntiMoji obtained an F1-score of 0.840, 0.861, 0.757, 0.811, and 0.829 respectively. This is also evident in the number of false positive (FP) utterances in the \emo{Joy} category, i.e., for the Stack Overflow dataset, the number of FPs for BERT, RoBERTa, ESEM-E, EMTk, and SEntiMoji are 34, 21, 29, 17, and 17 respectively, whereas, for flan-alpaca, ChatGPT, and GPT-4, the FPs are 259, 72, and 91.

Another common type of error was that the models predicted \emo{Neutral} often. \revisiontext{For instance, on the GitHub dataset GPT-4 identified 269 (67\%) utterances as \emo{Neutral}.}
In many cases a secondary or tertiary emotion for Shaver's categorization most closely describes the annotated utterances. However, those emotions were not provided to the model. For example, consider the following sentence from the GitHub dataset: \textit{``Any updates on this? I'm implementing a flutter application with barcode scanners, the soft keyboard on screen is really annoying."}, annotated as \emo{Anger} and, on a more granular level, as \emo{Annoyance}. All \zerollm models predicted it as \emo{Neutral}. 
As another example, the following sentence is annotated as \emo{Worry}, which is a tertiary-level emotion of \emo{Fear}: \textit{``My concern is that more new atributes may appear [...] it may break their behavior."}, while flan-alpaca and ChatGPT classified it as \emo{Neutral}.

We also observed a number of hallucinations in the \zerollm output~\cite{bang2023multitask}, where the models generated responses that were outside of what was asked. This led to situations where the models outputted emotions such as \emo{Apology} and \emo{Appreciation}, despite them not being in the prompted emotions list. For example, GPT-4 predicted the following sentence as \emo{Apology}: \textit{``Doh. Sorry for wasting your time."} even though the set of basic emotions provided in the prompt does not contain this emotion. 

In order to address these issues, we experiment with constructing prompts with a more granular level of emotions, i.e., by considering the second and tertiary-level emotions in Shaver's extended taxonomy.  This is also motivated by the study of Kocon et al.~\cite{kocon2023chatgpt}, who used all of GoEmotions' 27 emotions in their prompting experiments with ChatGPT.

\begin{table}[tb]
\centering
\small
\caption{Micro-averaged F1-score of emotion classification models for three different datasets.}
\begin{tabular} {|p{2cm}|c|c|c|} 
\hline
& GitHub~\cite{imran2022data} & SO~\cite{novielli2018gold} & JIRA~\cite{ortu2015bullies} \\ \hline
{\em SE-Specific} & & & \\ 
\quad ESEM-E & 0.440 & 0.674 & 0.744 \\ 
\quad EMTk & 0.434 & 0.651 & 0.734 \\ 
\quad SEntiMoji & 0.529 & 0.721 & 0.793 \\ 
\hline
{\em Fine-tuned} & & &\\ 

\quad BERT & 0.588 & 0.716 & 0.817 \\ 
\quad RoBERTa & \textbf{0.592} & \textbf{0.735}  & \textbf{0.818} \\ 

\hline

{\em Zero-shot} & &  & \\ 
\quad ChatGPT & 0.234 & 0.339 & 0.276 \\ 
\quad flan-alpaca & 0.424 & 0.293 & 0.432 \\ 
\quad GPT-4 & 0.355 & 0.444 & 0.256 \\ 

\hline

\end{tabular}
\label{tab:classification_summary}
\end{table}

\subsection{Granular-level Emotion Prompting}

In order to experiment with more granular emotions, we require a labeled dataset that includes these emotions. Therefore, we specifically conducted these experiments with Imran et al.~\cite{imran2022data}'s dataset, which provides a secondary and tertiary-level emotion annotation while the other datasets do not. 
First, we conducted prompt experiments using a part of Imran et al.'s training set (note that the \zerollm are not using the training data) varying the information used in the prompts for each instruct-tuned language model.
More specifically, we randomly selected 400 comments from the training dataset using stratified sampling and tested with granular-level prompting using the following strategies: 1) all emotions (basic, secondary and tertiary) from the extended Shaver's categories -- a total of 140 emotions; 2) only the basic and secondary emotions from the extended Shaver's categories -- a total of 34 emotions; 3) GoEmotions' list of 27 emotions.

We mapped the output emotion from the secondary and tertiary emotions to corresponding basic emotions as shown in Table~\ref{tab:shavers_category} and compared the results of the models at this level (as the SE-specific models can only produce results at the basic emotion level). 
We also found during the granular-level prompting that the models sometimes produced minor wording variations of the provided emotions, such as \emo{Confused} instead of \emo{Confusion}, \emo{Excited} instead of \emo{Excitement}. While mapping the outputs of the \zerollm to the basic emotions, we made adjustments as not to punish the models for these minor differences. 

During our experimentation, we observed that all three models tend to suffer more strongly from the issue of hallucination~\cite{bang2023multitask} when 
the complete emotion list (all of basic, secondary and tertiary level emotions) are provided. 
For example, GPT-4 suffered 50\% more hallucinations when the complete list is provided compared to the basic list of emotions. 
In particular, the models tended to generate extrinsic hallucinations~\cite{ji2022survey}, i.e., information beyond what is asked in the given prompt. This led to situations where the models generated emotions such as \emo{Concern}, \emo{Apology}, and \emo{Appreciation}, despite them not being in the prompted emotions. This suggests that providing a very large list of emotions may not be optimal. 

Out of the strategies we attempted, providing GoEmotions' 27 emotions list produced the best performance. For example, on the sample of the training dataset, ChatGPT achieved an F1-score of 0.201 when all emotions from Table~\ref{tab:shavers_category} were provided, 0.341 when basic and secondary emotions are provided, and 0.419 when GoEmotions' emotions are provided.
As noted earlier that the GoEmotions' taxonomy is developed specifically for text-based emotion recognition~\cite{goemotions}. This can explain why it performed better than emotions selected directly from Shaver's taxonomy, which was developed based on psychological evidence and not specifically for text~\cite{Shaver}.

Therefore, we opt to use GoEmotions' list of emotions for prompting for emotion classification using the \zerollm. Next, we report the results on the Imran et al.~\cite{imran2022data}'s held-out test dataset. 

\begin{table}[tb]
\centering
\scriptsize
\caption{Micro averaged F1-score of emotion classification for different models using Imran et al.'s dataset. The \zerollm use the GoEmotions list of 27 emotions.}
\begin{tabular} {|p{1.5cm}|c|c|c|c|c|c|c|} 
\hline
& \multicolumn{7}{c|}{Imran et al. (N=2000)~\cite{imran2022data}} \\ \cline{2-8}
& Anger & Love & Fear & Joy & Sad. & Surprise & Micro Avg.\\ \hline
{\em SE-Specific} & & & & & & & \\ 
\quad ESEM-E & 0.309 & 0.644 & 0.291 & 0.378 & 0.524 & 0.500 & 0.440 \\
\quad EMTk & 0.430 & 0.682 & 0.163 & 0.378 & 0.525 & 0.345 & 0.434 \\ 
\quad SEntiMoji & 0.460 & 0.642 & 0.377 & 0.556 & 0.629 & 0.458 & 0.529 \\ 
\hline
{\em Fine-tuned LLMs} & & & & & & & \\ 

\quad BERT & 0.506 & \textbf{0.712} & 0.536 & \textbf{0.579} & \textbf{0.636} & 0.594 & 0.588 \\
\quad RoBERTa & 0.525 &	0.683 & 0.492 & 0.500 & 0.613 & 0.673 & \textbf{0.592} \\

\hline

{\em Zero-shot LLMs} & & & & & & & \\ 
\quad ChatGPT & 0.337 &	0.492 & 0.182 &	0.458 & 0.417 & 0.511 & 0.429 \\
\quad flan-alpaca & 0.447 &	0.537 & 0.140 &	0.446 & 0.451 & \textbf{0.740} & 0.506 \\
\quad GPT-4 & 0.409 & 0.698 & 0.049 &	0.446 &	0.487 & 0.524 & 0.482 \\
 \hline
\end{tabular}
\label{tab:classification_results_emotions}
\end{table}

\noindent
{\bf Results and Discussion.} Table~\ref{tab:classification_results_emotions} shows the results for emotion classification on Imran et al.'s GitHub dataset for all the models. Overall, BERT and RoBERTa still achieve the best results with an average F1-score of 0.588 and 0.592 respectively, while among the SE-specific models deep learning-based SEntiMoji achieved the highest average F1-score of 0.529. From the table, it is clear that all three \zerollm improve in most categories of emotions and overall micro-averaged F1-score. It is also noticeable that they improved in distinguishing \emo{Love} and \emo{Joy} utterances. However, the \zerollm still perform badly for \emo{Fear}. Overall, surprisingly, the open-source model flan-alpaca achieved the best performance with an average F1-score of 0.506 -- an improvement of 19.33\% from the basic emotion-level prompting, while the proprietary model GPT-4 achieved 0.482 -- an improvement of 35.77\%. Both of these are improvement over the three SE-specific models and the proprietary ChatGPT ({\em gpt-3.5-turbo}) model. 

The results again point out that despite there having been major advancements in instruction-tuned LLMs, the fine-tuned deep learning models still perform better for specific, well-defined tasks that require domain-specific knowledge. To understand more where \zerollm are still making errors in detecting emotions, we conduct an error analysis on the errors in granular level prompting.

\begin{figure}[t]
\centering
\includegraphics[width=0.7\linewidth]{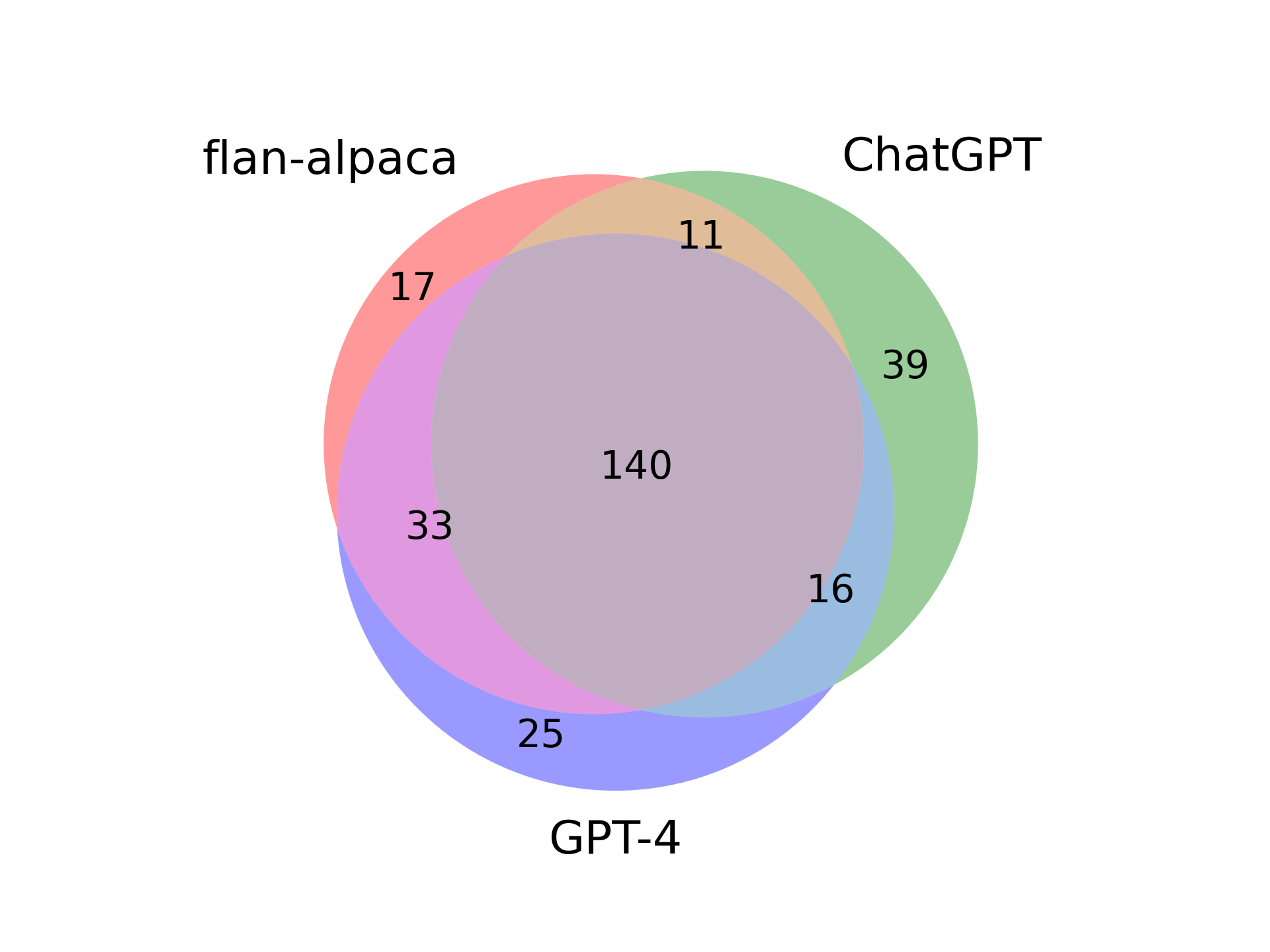}
\caption{Wrongly classified utterances among the \zerollm.
}
\label{fig:misclassification}
\end{figure}

\textbf{Error Analysis.} From the 356 non-\emo{neutral} instances in the test set, all three models correctly predicted 75 instances, two models made the right prediction for 67 instances, only one of the models made the right prediction for 74 instances, and no \zerollm made the right prediction on 140 instances. A Venn diagram of the classification error of each of the three \zerollm is shown in Figure~\ref{fig:misclassification}. We examine more closely the 140 utterances where all three \zerollm models made wrong predictions. 

As also noticeable in Table~\ref{tab:classification_results_emotions}, \emo{Fear} is the most often misclassified category with (35/140) instances. 
The errors in this category are especially discernible with GPT-4. With basic emotions only, GPT-4 achieved an F1-score of 0.353 in the \emo{Fear} category while at the granular level the F1-score went down to 0.049. The primary reason for it is that GPT-4 generated hallucinated output with labels such as \emo{Worry}, \emo{Concern}, etc., which are missing in the GoEmotions list. However, some of these emotions are present in Shaver's extended list and in Imran et al.'s annotation. In the annotated data, most of the \emo{Fear} utterances are due to the tertiary-level emotion \emo{Worry}.
For example, the utterance \textit{``Isn't this a breaking change? Can we get away with it?"} is annotated as \emo{Worry} (3rd level of {\em Fear}) in the ground truth. Another example is the utterance: \textit{``I guess my concern is that it sets a precedent where somebody could see it and think that it would be fine to use in `core'."} 

The second most misclassified emotion category is \emo{Joy} with (33/140) instances. Many of these errors are because the models are predicting conservatively, i.e., predicting \emo{Neutral} instead of a specific emotion. For example, \textit{``Anyway, the syntax change is fine."} -- this utterance is annotated as \emo{Approval} (2nd level of \emo{Joy}). Another example, \textit{``[USER] can you assign this ticket to me, I can help in this."} -- this utterance is annotated as \emo{Enthusiasm} (2nd level of \emo{Joy}). Also, notable here is that \emo{Enthusiasm} is not in GoEmotion's emotion list. Another type of error among \emo{Joy} category is that they are often misclassified as \emo{Surprise}. For example, \textit{``This was actually causing this test-case not to be executed!"} - this utterance is annotated as \emo{Relief} (2nd level of \emo{Joy}), but the flan-alpaca and GPT-4 model predicted as \emo{Surprise}. 

The third most misclassified category is \emo{Sadness} with (31/140) instances. We observed that these utterances are often misclassified as \emo{Anger}, \emo{Surprise}, or \emo{Neutral}. For example, flan-alpaca and GPT-4 predicted \emo{Surprise} for this utterance: \textit{``Ah sorry I thought `ScaleUpdateDetails' was constructed in `\_update' nvm."} 

We observed hallucinated emotions as well (\revisiontext{5 for GPT-4, 13 for flan-alpaca, 2 for ChatGPT}), especially \emo{Concern} and \emo{Worry} among \emo{Fear} utterances; and \emo{Appreciation} among \emo{Love} utterances. 

Overall, the error analysis points out the need for having a more specialized emotion taxonomy for text-based emotion detection, in particular for software-engineering-related text. As noted earlier, Shaver's~\cite{Shaver} taxonomy, developed in Psychology, includes many additional emotions that do not appear in the text and confuse the \zerollm. Meantime, while GoEmotions list focuses on text-based emotions, they are still missing some commonly observed emotions in software engineering such as \emo{Worry} and \emo{Frustration}. 
 \section{Emotion-Cause Extraction}
\label{emotion_cause_extraction}
The results of the preliminary study suggest that \zerollm are capable at detecting emotion categories when provided with granular level emotions, performing slightly worse than the best evaluated models (i.e., in Table~\ref{tab:classification_results_emotions}, flan-alpaca's F1-score is 0.506 relative to SEntiMoji's 0.529 and RoBERTa's 0.592).
In this section, we examine their feasibility for the more challenging task of \ece.

The use of LLMs for \ece has experienced a notable uptick in interest in recent years~\cite{poria2021recognizing, turcan-etal-2021-multi}. Emotion-cause extraction seeks to identify the cause or event that instigates a specific emotion in a given text, providing essential insights into human behavior and deepening our comprehension of the underlying emotions behind text-based communication. Researchers have explored the potential of LLMs in detecting emotion causes across multiple domains, such as social media and news articles~\cite{turcan-etal-2021-multi, li2021span, poria2021recognizing}.

Despite the growing interest in \ece in different domains, there is a lack of research on this problem in software engineering communication text. This research gap inspires our study, which aims to investigate the effectiveness of \zerollm in detecting emotion causes in GitHub comments.

To this end, we first manually annotate emotion causes in a subset of Imran et al.'s~\cite{imran2022data} data, identifying the text span that represents the cause of emotion in the comment. We then use zero-shot LLMs to extract emotion causes and compare their performance against the annotated emotion causes using the BLEU score~\cite{papineni2002bleu}, a standard metric in machine translation to evaluate text sequence similarity. Below, we present a detailed description of our annotation process, zero-shot LLMs, and the comparison of BLEU scores across different models and configurations.

\subsection{Annotation}

To create a dataset for the emotion-cause extraction task, we begin by selecting 75 utterances for each of the 6 basic emotion categories (\textit{Anger, Love, Fear, Joy, Sadness, Surprise}) from Imran et al.'s training dataset, totaling 450 utterances. Two senior undergraduate students (with 3+ years of experience in programming) are then tasked with annotating the dataset by identifying emotion causes, if any, based on the previously annotated basic, secondary, and tertiary emotions by Imran et al. We provide them with the following instructions:

\begin{quote}
{\em For each instance containing an emotion (Anger, Love, Fear, Joy, Sadness, Surprise), find the span of text (if any) that contributes to the annotated emotion. Each instance then should be annotated with its corresponding causes if existing. Emotion can sometimes be associated with more than one cause, in such a case, both causes should be marked. Since in some cases, more than one emotion can be present in an instance, the causes for emotion should be mapped as $<$emotion, cause span$>$.}
\end{quote}

The above instructions are adapted from Chen et al.'s seminal work on detecting emotion causes~\cite{chen-etal-2010-emotion}. We also provide the annotators with definitions and examples of different types of emotion causes. After the annotation task is completed, one of the authors of the paper manually reviewed both sets of annotations and noted disagreements in 44 of the 450 instances. To resolve these discrepancies, the annotators are asked to meet on Zoom and discuss and resolve their differences. This process ensures the annotated dataset's reliability and consistency.

\subsection{Model Selection}
For the automated \ece task, we evaluate the same three instruction-tuned models (ChatGPT, GPT-4, and flan-alpaca) that we used for emotion detection in Section \ref{prelim_study}, i.e., the preliminary study. We do not use BERT or RoBERTa as those models require a large amount of domain-specific training data~\cite{Lin2023SelectiveID}, which we lack. 

\subsection{Prompt Design}

The structure of our \ece prompt is intended to mimic a real-world scenario where a GitHub user is going through issues and pull requests, experiencing various emotions, and trying to pinpoint the cause of a specific emotion in a given utterance. We use a two-step prompt that asks the model to first detect the emotion in the utterance using the procedure outlined in Section \ref{prelim_study}. Then, we prompt the model to identify the cause of this emotion, as shown in the framed box structure.

\begin{tcolorbox}[breakable,colback=black!03!white]

\sf{
\noindent
You are a GitHub user. You are reading GitHub comments. Your task is to extract the span that is causing the emotion \textcolor{blue}{\em <insert emotion>} in the following GitHub utterance: \textcolor{blue}{\em <insert utterance>}. 

\smallskip
\noindent
Write the span of the cause within a double quote.

\smallskip
\noindent
Do not write anything else. 
}

\end{tcolorbox}

\subsection{Results} 

To ensure consistency in our evaluation, we preprocess all comments, annotated causes, and model-extracted causes by removing punctuation, lemmatizing, and stemming. After preprocessing, the average length of the 450 utterances is 28.08 words, while the average length of the manually annotated emotion cause spans is 7.43 words. We find that the emotion cause spans extracted by GPT-4, ChatGPT, and flan-alpaca have average lengths of 8.85, 8.64, and 13.12 words, respectively.

\subsubsection{BLEU score} The BLEU (Bilingual Evaluation Understudy) score is a metric used to evaluate the quality of machine-generated text by comparing it to human-generated reference text~\cite{papineni2002bleu}. The BLEU score measures the similarity between the machine-generated text and the reference text based on the n-gram overlap between them. The higher the BLEU score, the closer the machine-generated text is to the reference text. 
The formula for the BLEU score is: $BLEU = BP \cdot \exp\left(\sum_{n=1}^{N}w_n \cdot \log(p_n)\right)$. where,
\begin{itemize}
\item $BP$ is the brevity penalty, which is 1 if the machine-generated text is longer than the reference texts and less than or equal to them otherwise.
\item $N$ is the maximum n-gram order.
\item $p_n$ is the precision score for n-grams.
\item $w_n$ is the weight for n-grams, which is usually set to $\frac{1}{N}$ for uniform weighting of all n-gram orders.
\end{itemize}

\subsubsection{BLEU Score Interpretation} 

The interpretation of BLEU scores can vary depending on the specific domain and language being evaluated.  In the software engineering domain, a BLEU score is commonly used to evaluate the quality of generated bug reports, code comments, and code summarization. Denkowski and Lavie~\cite{denkowski2010choosing} suggest that BLEU scores above 0.30 generally indicate that the generated text is understandable, while scores above 0.50 are indicative of good and fluent results. Previous research~\cite{Seljan2012BLEUEO, denkowski2010choosing}, including studies in software engineering~\cite{gao2020generating}, has used this scale to interpret the results of BLEU scores. It is important to note that the choice of n-gram order used to calculate the BLEU score can impact the final score; typically, 4-gram is used for BLEU score calculation~\cite{denkowski2010choosing, papineni2002bleu}. In the case of our study, however, the emotion cause spans are often short, making the bigram a more suitable choice for BLEU score calculation, i.e., BLEU-2.

\subsubsection{Discussion} 

The BLEU scores for the three models using unigram, bigram, trigram, and four-gram are shown in Table~\ref{tab:bleu_score}. The score ranges between 0.450 to 0.637, which indicates that all models are generally able to extract the right emotion causes to some extent, especially GPT-4 and flan-alpaca as both models' BLEU scores are always above 0.5.

\begin{table}[tb]
\centering
\caption{BLEU scores of different \zerollm.}
\small
\begin{tabular} {|l|c|c|c|c|}
\hline
Model & BLEU-1 &  \textit{BLEU-2} & BLEU-3 & BLEU-4 \\ \hline\hline
ChatGPT & 0.522 & \textit{0.489} & 0.467 & 0.450 \\
GPT-4 & {\bf 0.637} & {\bf \textit{0.598}} & {\bf 0.571} & {\bf 0.554} \\
flan-alpaca & 0.571 & \textit{0.543} & 0.525 & 0.508 \\
\hline
\end{tabular}
\label{tab:bleu_score}
\end{table}

When considering BLEU-2, GPT-4 obtains the highest score of 0.598, followed by flan-alpaca with 0.543 and ChatGPT with 0.489. Out of 450 utterances, 107 cases are identified where all three models' BLEU-2 scores are higher than 0.5. We observe that these 107 utterances are relatively short, with an average length of 15.26 words, while the annotated cause spans have an average length of 7.02 words. The three models, GPT-4, ChatGPT, and flan-alpaca, extract similar length spans on average, which are 7.89, 7.79, and 7.95 words, respectively. For example, in the following utterance, \textit{``I'm not sure how to fix this, nor if this is acceptable in this test case. Namespaces in TS are magic to me ${grinning-face-with-sweat}$"}, the annotated cause of \emo{Amusement} (3rd level \emo{Joy}) is \textit{``Namespaces in TS are magic to me"}. GPT-4 also extracted the same span as the cause. However, it is not always the case that the annotated cause span completely overlaps with the spans extracted by the models. For example, in this utterance, \textit{``Oh, you didn't add composes and values. Well, I like it even more. Those features are hard to maintain."}, the annotated cause span is \textit{``I like it even more"}, and the extracted cause span by GPT-4 is \textit{``Well, I like it even more."}

Out of 450 utterances, we observe that in 41 cases, all three models' BLEU scores are less than 0.30. These comments are relatively longer, containing an average of 44.17 words, while the annotated cause spans contain an average of 5.05 words. The extracted average lengths of spans for GPT-4, ChatGPT, and flan-alpaca are 10.10, 13.14, and 22.83 words, respectively. 

\subsubsection{Error Analysis} 

To gain insight into the models' mistakes, we analyze the 41 utterances where all three models had a BLEU score of less than 0.30. Our examination reveals that the errors can be classified into a few primary categories, which are elaborated below.

\noindent
\textbf{Incorrect Emotion.} The main source of error for all three models is the misidentification of the emotion expressed in the utterance (24/41 utterances). This misidentification leads to the detection of an incorrect cause event. For instance, consider the utterance, ``\textit{Oh right! ${upside-down-face}$ This started as a Mac issue, I forgot to add the rest.}" The annotated emotion for this utterance was \emo{Neglect} (2nd level \emo{Sadness}) and the annotated cause span is ``\textit{I forgot to add the rest.}" However, ChatGPT identifies the utterance as \emo{Confusion} (2nd level \emo{Surprise}) and extracts ``\textit{${upside-down-face}$}" as the cause event instead. GPT-4 detects \emo{Amusement} in the utterance and extracts the cause span as \textit{``Oh right! ${upside-down-face}$."} Meanwhile, flan-alpaca identifies ``\emo{Curiosity} (2nd level \emo{Surprise})" and extracts the cause span as \textit{``Oh right!"}
This error category emphasizes the importance of accurately detecting the emotion expressed in the text before extracting emotion causes.

\noindent
\textbf{Incorrect Cause.} This error occurs when the models correctly classify the emotion but detect a different cause than the ground truth (12/41 utterances). For example, in the following utterance \textit{``[USER] yep, it is bug, we will fix it, so we have it in `experiments` :+1:"}, the annotated emotion is \emo{Approval}, and the annotated cause span is \textit{``it is a bug"}, while GPT-4 detected the cause span \textit{``we will fix it"}. This error category highlights the difficulty in identifying the exact cause of events in conversational text, especially in longer, multi-part comments.

\noindent
\textbf{Hallucinations.} 
In addition to the two error categories described above, we also observe instances of hallucinations in the cause event extraction process. In some cases, the models' output \textit{``the entire sentence."}, \textit{``the span: $<$followed by the span$>$"},  \textit{``span starting from word X to word Y"}, and other nonsensical outputs. We observe that ChatGPT produces more hallucinated data than the other two models, which is one reason why its BLEU score is lower. This highlights the need for continued research into developing more accurate and reliable models that can follow the prompt exactly.
 \section{Investigating the Causes of Frustration in the Tensorflow Repository: A Case Study}

\emo{Frustration} is a pervasive emotion in software development~\cite{ford2015exploring}, and it is particularly relevant in the context of open-source projects~\cite{morgan2014lessons}. Wrobel et al. noted that \emo{Frustration} is the most commonly felt emotion during software development~\cite{wrobel2013emotions}. Collaborative work, lack of control over external contributors' code, and the complexity of software development processes can all contribute to the \emo{frustration} of developers and end-users. In contrast to other emotions, such as \emo{confusion} or \emo{excitement}, \emo{frustration} is more strongly associated with obstacles, challenges, and difficulties. It is also often accompanied by other negative emotions, such as \emo{anger}, \emo{disappointment}, or \emo{helplessness}~\cite{Gelbrich2010AngerFA}. Given the complexity and collaborative nature of open-source software development, \emo{frustration} is undesirable but likely to be a common experience for many contributors and users. Therefore, understanding the causes of \emo{frustration} in open-source development can provide valuable insights for project maintainers into what are the key issues that impede collaboration and the productivity of project participants.

Tensorflow\footnote{\url{https://github.com/tensorflow/tensorflow}} is a popular open-source platform for developing machine learning models and has a large number of developers and a huge user-base, which makes it an interesting case study for investigating the causes of \emo{frustration} in open-source software development. 
For instance, monitoring of the causes of \emo{frustration} in TensorFlow contributors can aid in the construction of project maintainer dashboards that help attract and retain open source contributors~\cite{storey2019software,oss-dash}.

\subsection{Data Collection and Cause Extraction}

To conduct our analysis, we collect all publicly available issues and pull requests comments made on the Tensorflow repository, hosted on GitHub, between March 30, 2022, and March 30, 2023. We choose this time period to ensure that our analysis covers a recent and substantial range of comments. 
Most GitHub repositories, including Tensorflow, differentiate different types of comment authors based on their relationship to the project, such as Contributors, Collaborators, Members and None\footnote{\url{https://docs.github.com/en/graphql/reference/enums\#commentauthorassociation}}.
A Collaborator is a GitHub user invited to work on the repository, a Contributor has committed code before, a Member belongs to the owning organization, and None has no affiliation with the repository. Collaborators, Contributors, and Members are active developers, while None comprises user commenters. To analyze software developer \emo{frustration}, we exclude comments from the None category.

Following the \ece procedure described in Section \ref{emotion_cause_extraction}, we extract the emotions and causes of each comment. We use the \textit{flan-alpaca} model for this purpose, as it performed reasonably well in both emotion detection and \ece tasks compared to the proprietary \zerollm. Another advantage of \textit{flan-alpaca} is that it is open-source and its weights are publicly available. This ensures the reproducability of our results. In contrast, closed-source LLMs may become unavailable, e.g., OpenAI's Codex LLM was deprecated in March, 2023.

We collect only the utterances that the model identified as expressing \emo{Frustration}, resulting in a dataset of 1275 comments.

\subsection{Clustering}

To identify common themes among the causes of \emo{frustration}, we employ the DBSCAN (Density-Based Spatial Clustering of Applications with Noise) algorithm~\cite{ester1996density}. It has been effectively used in previous software engineering studies involving clustering textual data~\cite{villarroel2016release, scalabrino2017listening}.
The main advantage of using the DBSCAN algorithm is that it does not require a pre-specified number of clusters, which can be difficult to estimate in advance. This is particularly useful in the context of identifying common themes among the causes of \emo{frustration}, as it is difficult to know beforehand what the common themes are. Another advantage of the DBSCAN algorithm is its ability to automatically handle noise and outliers, which is relevant as the extracted causes by \textit{flan-alpaca} can contain errors, as discussed in the previous sections.

While DBSCAN does not require to specify the number of clusters, it requires two key parameters~\cite{schubert2017dbscan}: 1) $\epsilon$ - a real positive value - the maximum allowed distance between two samples to be considered that they are part of the same dense region, and 2) $MinPts$ - a small positive constant integer - the minimum number of samples required to consider a dense region as a cluster. We performed a manual parameter sweep, testing $\epsilon$ values from 0.1 to 0.8 in increments of 0.05, and $MinPts$ values from 2 to 6, following standard guidelines for parameter tuning in machine learning and data mining~\cite{hastie2009elements}. Based on the number of clusters, average number of elements per cluster, and cluster composition, we selected $\epsilon$ = 0.3 and $MinPts$ = 4, which yielded 23 clusters. Before applying the DBSCAN algorithm, we perform standard text pre-processing such as removing punctuation, URL removal, and lemmatizing on the list of causes. We use the {\sf scikit-learn} library's implementation of the DBSCAN algorithm with cosine similarity and sentence-level embeddings (\textit{all-mpnet-base-v2} model~\cite{reimers2019sentence}). 

To focus our analysis on the most common causes of \emo{frustration}, we limit our discussion to the top 6 clusters in terms of the number of comments in each cluster. The clusters are presented in Table~\ref{tab:clusters}, along with their description, size, and examples. We read the GitHub comments and the emotion causes to identify the underlying theme in each cluster that leads to \emo{frustration}.

\begin{table*}[t]
\centering
\small	
\caption{Clusters of causes of \emo{Frustration} in TensorFlow project participants in GitHub.}
\vspace{-0.2cm}
\begin{tabular} {|l|p{7.5cm}|c|p{8cm}|}
\hline
& Cluster Description & Count & Example Comments \\ \hline

\multirow{5}{*}{1}
& \textbf{TensorFlow Version and Dependency Issues}: This cluster focuses on build and compatibility problems across various TensorFlow versions, challenges in reproducing issues in specific TensorFlow versions, and complications with related libraries and plugins such as TensorRT and Keras.
& 58
& (1) {\em [USER] Your original issue looks like \textbf{you have a bad version of tensorflow\_io\_gcs\_filesystem installed}. [...].} \newline
(2) {\em \textbf{It's probably not a bug in Tensorflow but Apple's tensorflow metal plugin}. See for example the following discussion [...]} 
\\ \hline

\multirow{4}{*}{2}
& \textbf{Pull Request Delays and Merge Conflicts}: The cluster comprises developer frustration from unresolved merge conflicts and from delays in merging pull requests.
& 26
& (1) {\em [...] But \textbf{there are a bunch of merge conflicts}. Since Random seeds are such a common topic in software [...]} \newline
(2) {\em It might have been a \textbf{wrong-way merge} or something like that. At this point it's usually easier to just close it [...]}
\\ \hline

\multirow{4}{*}{3}
& \textbf{Failing Tests}: This type of \emo{frustration} arises from the ambiguity and complexity of test failures, which make it challenging for project participants to determine whether the issues are linked to their code changes or are caused by unrelated factors.
& 15
& (1) {\em [USER]: It is just a first draft. \textbf{The test doesn't even work}. In the meantime, [...]}\newline
(2) {\em [...] Yes, I'll work on this. \textbf{It's weird that these tests are failing because I thought I ran them successfully for PR} [...]}
\\ \hline

\multirow{4}{*}{4}
& \textbf{Too Fine-Grained Commits}: The cluster reflects developer \emo{frustration} caused by too granular commits in the repository. Some developers request a commit history devoid of incremental commits that represent only partial progress on a change task.
& 9
& (1) {\em Can you squash these commits please? \textbf{It doesn't make sense to have 5 commits for a line change} and one extra empty line.} \newline
(2) {\em \textbf{3 commits for a single line change}? Can you please merge the commits in just one? [...]}
\\ \hline

\multirow{3}{*}{5}
& \textbf{CI Flakiness}: This type of \emo{frustration} is caused by Continuous Integration (CI) failures that seem unrelated, inconsistent, or uninformative to developers. 
& 8
& (1) {\em [USER] \textbf{there was failed ci.} Is there anything to do?}\newline
(2) {\em \textbf{CI failure does not look related to these changes}, seeing the same failure on \#56345 [...] so I assume this is noise. [...]}
\\ \hline

\multirow{4}{*}{6}
& \textbf{CUDA/CuDNN Compatibility Issues}: This cluster reflects the \emo{frustration} experienced when dealing with compatibility issues related to CUDA and CuDNN. 
& 8
& (1) {\em Unfortunately this change needs to be rolled back, \textbf{it seems it breaks JAX build under CUDA 11.4 and CuDNN 8.2}}\newline
(2) {\em [...] - Did you downgrade the CUDA to 11.2? \textbf{Looking at Nvidia docs it looks like the display driver and cuda driver do not match} [...]}
\\ \hline

\end{tabular}
\label{tab:clusters}
\end{table*}

\subsection{Causes of Frustration} 
We utilized \textit{thematic analysis} to identify the themes of the clusters~\cite{maguire2017doing}. Specifically, one of the authors of this paper read each comment and coded the initial themes. Then another author reviewed the themes, then both authors discussed resolving discrepancies and finalizing the themes until the analysis reached saturation, with no new themes emerging~\cite{saunders2018saturation}.
Each cluster theme is described below:

\smallskip
\noindent
\textbf{TensorFlow Version and Dependency Issues}: 
This cluster primarily includes project participants struggling with incompatibility issues due to version mismatches between TensorFlow and its related dependencies. They express frustration over difficulties in configuring TensorFlow to operate correctly on their system. They also express frustration over transitioning from legacy versions to newer versions. One possible way to address these issues is to provide a more comprehensive documentation on version compatibility between TensorFlow and its dependencies.

\smallskip
\noindent
\textbf{Pull Request (PR) Delays and Merge Conflicts}: This cluster is related to PR merging and associated communication, as well as merge conflicts. The project participants express \emo{frustration} when they have to wait a long time for a PR to be reviewed or merged by the project maintainers. Merge conflict-induced \emo{frustration} is a well-known issue in open source software development~\cite{kudrjavets2022mining}. Implementing automated review bots and streamlined conflict-resolution procedures can help mitigate this form of \emo{frustration}.

\smallskip
\noindent
\textbf{Failing Tests}: The cluster highlights the \emo{Frustration} felt due to test failing, possibly flaky tests~\cite{parry2022surveying}. The project participants report two main sources of \emo{frustration}: first, the inability to identify the root cause of test failures that seem unrelated to their code changes; second, unexpected test failures leading to their PRs being reverted.

\smallskip
\noindent
\textbf{Too Fine-Grained Commits}: This cluster reflects developers' \emo{frustration} on commits that capture incomplete changes or partial progress on a task, which need to be squashed. The comments demonstrate developer sensitivities around balancing incremental changes with maintaining a coherent commit history. Setting PR guidelines about git commit hygiene can help to mitigate this issue.

\smallskip
\noindent
\textbf{CI Flakiness}: Like test flakiness, CI flakiness is another common source of developer \emo{Frustration}~\cite{widder2019conceptual, parry2022surveying}. This cluster highlights the complexity of CI failures. The \emo{frustration} is evident as the developers grapple with failed CI tests, yet believe these problems are unrelated to their own contributions.

\smallskip
\noindent
\textbf{CUDA/CuDNN Compatibility Issues}: The project participants express \emo{frustration} regarding GPU library compatibility. This reflects the challenge of managing interdependent, rapidly evolving software ecosystems~\cite{berman2023machine}. TensorFlow relies on quickly changing GPU libraries like CUDA and CuDNN. Expanding CI testing across more diverse versions, detecting CUDA/CuDNN versions and alerting if incompatible, and explicitly documenting supported versions can help to reduce this pitfall. \section{Related Work}

The related work can be divided into three parts: prompt engineering for \zerollm, automated \ece in NLP, and the role of emotions in software engineering.

\smallskip
\noindent
\textbf{Prompt Engineering for Zero-Shot LLMs.}  Zero-shot learning, a task where a model is trained to recognize and classify unseen classes without any explicit training data for those classes, has been a recent focus among researchers and practitioners for a variety of tasks, including image and text classification, question answering, language generation, and data augmentation~\cite{li2021prefix, lazaridou2022internet, zhou2022learning, Lin2023SelectiveID}. Recently, researchers have focused on leveraging LLMs for zero-shot learning~\cite{xian2018zero, wei2021finetuned, zhong2021adapting, zhang2017learning}. In the context of zero-shot learning, prompt engineering with LLMs has emerged as an area of interest in recent years~\cite{zhong2021adapting, wei2021finetuned, kojima2022large, Lin2022ZeroShotRD}. One approach that has been explored is the use of task-specific prompts, which are designed to elicit the desired response from the model. These prompts can be constructed manually or generated automatically and can be tailored to the specific task at hand~\cite{kojima2022large}. For example, Brown et al. used an LLM to perform text classification using task-specific prompts~\cite{brown2020language}. Another approach is the use of general-purpose prompts, which are designed to be broadly applicable across a range of tasks~\cite{zhong2021adapting, OpenAI2023GPT4TR}. The recent advancements in language models such as ChatGPT~\cite{chatgpt}, GPT-4~\cite{OpenAI2023GPT4TR}, BARD~\cite{bard}, LLaMA~\cite{llama}, and Alpaca~\cite{alpaca} have made the general-purpose prompt approach increasingly popular. These models have achieved impressive performance across a range of tasks and continue to push the boundaries of NLP.

\smallskip
\noindent
\textbf{Automated Emotion-Cause Extraction in NLP.} Automatically extracting emotion-cause has gained attention in recent years in NLP~\cite{turcan-etal-2021-multi, li2021span, poria2021recognizing, zhang2022cl, xu2021two, xia2019emotion}. Emotion-cause extraction is challenging, as both emotions and their causes can be expressed in various ways, including but not limited to explicit statements, implicit suggestions, and contextual cues. Several techniques have been proposed to address this challenge, including rule-based approaches, machine learning-based approaches, deep learning-based approaches, and LLM approaches~\cite{lee2010text, gui2018event, gui2017question, poria2021recognizing}. In recent years, the focus has been on LLM approaches~\cite{turcan-etal-2021-multi, li2021span, poria2021recognizing}. Researchers have explored this area with prompting as well~\cite{zheng2022ueca}. Wang et al. noted that ChatGPT achieves comparable performance on the emotion-cause extraction task in news articles~\cite{wang2023chatgpt}. In this study, we apply prompt-based emotion-cause extraction for three state-of-the-art LLMs, namely ChatGPT, GPT-4, and flan-alpaca~\cite{chatgpt, OpenAI2023GPT4TR, flan}.

\smallskip
\noindent
\textbf{The Role of Emotions in Software Engineering.} Emotions play a crucial role in software engineering, as software development is a complex and collaborative process that often involves multiple stakeholders with different perspectives and priorities~\cite{Murgia2014DoDF, girardi2021emotions, graziotin2015understanding, madampe2023framework, wurzel2022interpersonal}. 
Researchers have explored the role of emotions in software engineering through qualitative analyses, quantitative analyses, and surveys~\cite{Sajadi2023, madampe2023framework, wrobel2013emotions, muller2015stuck, graziotin2017unhappiness, graziotin2018happens, Murgia2014DoDF, esem-e, emtk, sentimoji, egelmanPredicting2020}. Gachechiladze et al. conducted a study on where \emo{Anger} is directed, i.e., towards self, others, and objects~\cite{gachechiladze2017anger}. Ford et al. conducted a survey with 256 software developers to identify common sources of \emo{frustration}~\cite{ford2015exploring}. Graziotin et al. investigated the causes of unhappiness among software developers, using a survey of 2,220 participants~\cite{graziotin2017unhappiness}. Later, Graziotin et al. conducted a study of the effects of unhappiness~\cite{graziotin2018happens}. More recently, there has also been a focus on studying conflicts, toxicity, and incivility in open source communities~\cite{incivility, miller2022did, wurzel2022interpersonal, Ehsani2023_FSE, sarkerAutomated2022}.  

To our best knowledge, there has been no research on the automated detection of emotion-causes in software engineering. To fill this gap, in this study, we examine the efficacy of existing state-of-the-art large language models in automatically extracting emotion-causes. We also perform a case study to demonstrate how these models can be applied in real-world scenarios.
 \section{Threats to Validity} In this section, we discuss the potential threats to the validity of our study grouped into three categories: construct validity, internal validity, and external validity.

\smallskip
\noindent{\textbf{Construct validity.}} Construct validity is the extent to which our study accurately measures the concepts and constructs it aims to measure. One potential threat to construct validity is the use of automated \zerollm to extract emotion causes from domain-specific comments. These models are designed to perform general-purpose tasks and are not fine-tuned to extract emotion causes in software engineering communication text. To address this threat, we perform multiple error analyses to understand where these models make mistakes. Additionally, there could be a threat in the construction of the prompts. To mitigate this threat we followed existing literature and validated various versions of the prompt with labeled data in order to find a suitable prompt for the \zerollm. Another threat to construct validity comes from our manual labeling of the causes, which may introduce some subjectivity and bias, potentially impacting the accuracy of the reported results. We reduced this threat via multiple annotators and by resolving discrepancies to achieve 100\% agreements.

\smallskip
\noindent{\textbf{Internal validity.}} The concept of internal validity relates to the degree to which the manipulation of an independent variable is responsible for the outcomes of a study. In our examination of an open-source project, \emo{frustration} causes represent an independent variable. However, there are potential threats to internal validity, such as unaccounted factors like prior experience with the project or technical expertise that could contribute to software developers' \emo{frustration}. Moreover, the use of flan-alpaca for extracting frustration causes could result in the misclassification of some utterances, leading to the potential omission of certain clusters that could provide alternative explanations for \emo{frustration} or identification of some clusters that do not in fact represent this emotion. Nevertheless, the use of DBSCAN reduces the effect of random noise, and the list of \emo{frustration} causes provided in Table~\ref{tab:clusters} follow the software engineering literature on common problems developers face during open-source software development~\cite{kinsman2021software, li2021code, gousios2016work}.

\smallskip
\noindent{\textbf{External validity.}} External validity pertains to the generalization of our study's findings to other settings and contexts. For emotion detection, we used the categories from extended Shaver's taxonomy as well as GoEmotions' taxonomy from previous research~\cite{imran2022data, goemotions, Shaver}.
However, our findings may not necessarily be transferable to other emotion categories. Another potential threat to external validity is the specific nature of the open-source project we studied, i.e., TensorFlow. The project's characteristics, such as its size, development stage, and community culture, may not be representative of other open-source projects. Additionally, the programming language and technology stack used in the project may have influenced the types of causes of  \emo{frustration} observed. Therefore, it is important to interpret our findings in the context of the specific project we studied and exercise caution when generalizing them to other open-source projects. Further investigation is needed to generalize these results beyond the three specific models and the data and projects we have used in our study. 

\section{Conclusions}

In this paper, we presented an approach for automated \ece in software developer communication using three \zerollm, namely ChatGPT, GPT-4, and (the open-source) flan-alpaca, through a prompting approach. We first conducted a preliminary study to evaluate the models' performance in emotion classification tasks on an existing recent dataset, and we found that they perform well compared to state-of-the-art models. We then showed the feasibility of using these models for emotion-cause extraction on a subset of 450 utterances from the same dataset by manually annotating the emotion causes of these utterances and automatically extracting the causes using prompts. We compared the BLEU score performances of the models and found that GPT-4 achieved the highest BLEU-2 score of 0.598, followed by flan-alpaca with 0.543, and ChatGPT with 0.489. To demonstrate the possible real-world applications of \ece, we conducted a case study on the causes of \emo{frustration} in a large GitHub open-source project -- Tensorflow. 

There are several avenues for future work. First, our case study only focused on one emotion and one open-source project. Future studies that use \ece should investigate other emotions and a broader range of projects to generalize our findings. Second, further work is needed to improve the accuracy of \ece from text in software engineering communication. This could involve few-shot prompting, fine-tuning language models, or developing domain-specific models tailored for software engineering communication. Overall, our study provides a starting point for future research to explore the potential of \ece in software engineering communication.

\bibliographystyle{IEEEtran}
\bibliography{references}

\end{document}